\documentclass[preprintnumbers, prd, twocolumn, floatfix,
preprintnumbers,nofootinbib,showpacs,
letterpaper, amsmath, amssymb,  superscriptaddress]
{revtex4}
\usepackage{graphicx}
\usepackage{epsfig}
\usepackage{bm}
\usepackage{amsfonts}
\usepackage{epstopdf}
\usepackage[latin1]{inputenc}
\usepackage{graphicx}
\usepackage{amssymb}
\usepackage{color}
\usepackage{float}
\usepackage{amsmath}
\usepackage{amsfonts}
\usepackage{dcolumn}
\usepackage{hyperref}
\usepackage{epsfig}
\usepackage{bm}

\def\be{\begin{equation}}
\def\ee{\end{equation}}
\def\ba{\begin{eqnarray}}
\def\ea{\end{eqnarray}}

\begin{document}

\title{$F(R)$ nonlinear massive theories of gravity and their cosmological implications}

\author{Yi-Fu Cai}
\email[]{yifucai@physics.mcgill.ca}
\affiliation{Department of Physics, McGill University, Montr\'eal, QC H3A 2T8, Canada}

\author{Francis Duplessis}
\email[]{francis.duplessis@mail.mcgill.ca}
\affiliation{Department of Physics, McGill University, Montr\'eal, QC H3A 2T8, Canada}
\affiliation{Department of Physics, Arizona State University, Tempe, AZ 85287, United States}

\author{Emmanuel N. Saridakis}
\email[]{Emmanuel\_Saridakis@baylor.edu}
\affiliation{Physics Division, National Technical University of Athens, 15780 Zografou Campus, Athens, Greece}
\affiliation{Instituto de F\'{\i}sica, Pontificia Universidad de Cat\'olica de Valpara\'{\i}so, Casilla 4950, Valpara\'{\i}so, Chile}

\begin{abstract}
We propose a nonlinear massive gravitational theory which includes $F(R)$ modifications. This construction inherits the benefits of the de Rham-Gabadadze-Tolley model and is free of the Boulware-Deser ghost due to the existence of a Hamiltonian constraint accompanied by a nontrivial secondary one. The scalar perturbations in a cosmological background can be stabilized at the linear level for a wide class of the $F(R)$ models. The linear scalar mode arisen from the $F(R)$ sector can absorb the
nonlinear longitudinal graviton, and hence, our scenario demonstrates the possibility of a gravitational Goldstone theorem. Finally, due to the combined contribution of the $F(R)$ and graviton-mass sectors, the proposed theory allows for a large class of cosmological evolutions, such as the simultaneous and unified description of inflation and late-time acceleration.
\end{abstract}


\maketitle

\section{Introduction}

The search for a consistent theory of massive gravity has been open for
decades. Its motivations arise from both theoretical considerations, namely
to understand the construction procedure of a massive spin-2 theory, as well
as (lately) from observational requirements, that is to explain the universe
acceleration through such an Infra-Red (IR) modification of General
Relativity. However, since the first, linear approach \cite{Fierz:1939ix},
the  subject remains a theoretically intriguing problem.

In the instructive idea of Fierz and Pauli  \cite{Fierz:1939ix} General
Relativity is extended by introducing a linear mass term and thus the theory
involves at least 5 degrees of freedom (DoF), representing a massive spin-2
field in a Poincar\'e invariant background. However, it turns out that the
graviton's longitudinal DoF remains coupled to the trace of the
energy-momentum tensor regardless the smallness of the graviton mass. This
leads to the famous van Dam-Veltman-Zakharov (vDVZ) discontinuity
\cite{vanDam:1970vg} and thus to a severe challenge by experiments and
observations. This discontinuity can be alleviated at the nonlinear level
through the Vainshtein mechanism \cite{Vainshtein:1972sx}, however, due to
the constraints on the dynamical variables, the same nonlinearities give rise
to a ghost instability, called Boulware-Deser (BD) ghost
\cite{Boulware:1973my}. Using the effective field theory approach one can
show that the BD ghost is related to the Goldstone boson associated with the
broken
general covariance \cite{ArkaniHamed:2002sp}.

The above inconsistencies puzzled physicists for years. Recently de Rham,
Gabadadze and Tolley (dRGT) showed that the BD ghost can be removed in a
suitable nonlinear massive gravitational theory \cite{deRham:2010ik}. In
particular, due to a delicate construction of the graviton potential, the
Hamiltonian constraint and the associated secondary one are restored, and
thus this IR modified theory becomes free from BD ghosts
\cite{Hassan:2011ea}. Apart from the theoretical interest, dRGT construction
has the additional advantage that its application to a cosmological framework
leads to late-time cosmic acceleration, where a sufficiently small value of
the graviton mass mimics an effective cosmological constant
\cite{deRham:2010tw, Gumrukcuoglu:2011ew, Comelli:2011zm}.
However, as was shown in \cite{DeFelice:2012mx}, cosmological perturbations
of the dRGT massive gravity around background solutions exhibit
instabilities.

On the other hand, after the 60's physicists realized that although General
Relativity is not renormalizable, possible high-energy corrections could
improve renormalizability and thus quantization \cite{Utiyama:1962sn,
Stelle:1976gc}. Although these Ultra-violet (UV) corrections are expected to
be of quantum origin or arise from an underlying fundamental theory such as
string theory (for example see \cite{Deser:1977nt, Vilkovisky:1992pb}), one
can describe them effectively, by investigating a classical, modified,
gravitational action. The simplest model of such an UV modified gravity, that
can sufficiently encapsulate the basic properties of higher-order
gravitational theories, is the $F(R)$ paradigm, in which the gravitational
Lagrangian is extended to an arbitrary function of the Ricci scalar (see
\cite{DeFelice:2010aj} for a review). The corresponding $F(R)$ cosmology is
able to describe the inflationary epoch, and in particular the well-known
Starobinsky's $R^2$-inflation scenario \cite{Starobinsky:1980te} proves to be
the best-fitted scenario with the recently-released Planck data
\cite{Ade:2013uln}.

Inspired by the above discussion, in this Letter we propose a modification of
General Relativity both in the UV and IR regimes, that is the $F(R)$
nonlinear massive gravity. In this theory, the extra scalar DoF of the $F(R)$
sector, clearly seen through a conformal transformation, has a
positive-defined kinetic term as usual, and its interaction with the massive
sector can stabilize metric perturbation of scalar type at linear order
(this is a novel feature, not present in \cite{Nojiri:2012re}). In summary,
the total theory is not only free of BD ghosts at the fundamental level, but
it is also free of linear cosmological perturbative instabilities for the
largest part of its parameter space, even in homogeneous and isotropic
geometries. Finally, the increased freedom of both $F(R)$ and
massive-graviton sectors can lead to a large class of interesting
cosmological behaviors at early and late times, in agreement with
observations.

\section{The setup}

Imposing both the UV ($F(R)$ sector) and IR (graviton-mass sector)
modifications, the total action becomes
\begin{eqnarray}\label{action}
 S = \frac{M_p^2}{2} \int d^4x \sqrt{|g|} ~ [ F(R) + 2m_g^2 {\cal U}_M ] ~,
\end{eqnarray}
where $M_p$ the Planck mass, $g$ the physical metric and $m_g$ the graviton
mass. As usual in dRGT construction, to build the dimensionless graviton
potential ${\cal U}_M$  one needs to define the matrix $\mathbb{K} \equiv
\mathbb{I} - \mathbb{X}$, where $\mathbb{X}\equiv \sqrt{g^{-1}f}$ involves a
non-dynamical (fiducial) metric $f$ \footnote{We use small $f$ to denote the
fiducial metric, and capital $F$ to denote the function $F(R)$.}. Then, the
regular anti-symmetrization in 4D space-time yields the following polynomials
\begin{eqnarray}\label{UMs}
 {\cal U}_2 = \mathbb{K}^\mu_{[\mu} \mathbb{K}^\nu_{\nu]} ~,~
 {\cal U}_3 = \mathbb{K}^\mu_{[\mu} \mathbb{K}^\nu_{\nu}
\mathbb{K}^\sigma_{\sigma]} ~,~
 {\cal U}_4 = \mathbb{K}^\mu_{[\mu} \mathbb{K}^\nu_{\nu}
\mathbb{K}^\sigma_{\sigma} \mathbb{K}^\rho_{\rho]} ~,
\end{eqnarray}
and the graviton potential is given by ${\cal U}_M = {\cal U}_2 + \alpha_3
{\cal U}_3 + \alpha_4 {\cal U}_4 $, containing two dimensionless parameters
($\alpha_3$, $\alpha_4$).

The UV sector inherits the remarkable properties of the $F(R)$ term. In
particular, by performing the conformal transformation $g_{\mu\nu}
\rightarrow \tilde{g}_{\mu\nu} = \Omega^2 g_{\mu\nu}$ with $\Omega =
\exp[\frac{\varphi}{\sqrt{6}M_p}]$, the $F(R)$ part can be reformulated as
the standard General Relativity minimally coupled to a canonical scalar field
$\varphi$, with effective potential
\begin{align}\label{potential0}
  U(\varphi) = M_p^2(RF_{,R}-F)/2F_{,R}^2~,
\end{align}
where $F_{,R}\equiv \partial{F}/\partial{R}$. Additionally, the conformal
transformation acts on the IR sector too, with the graviton potentials
transforming as
\begin{align}\label{potential}
 \tilde{\cal U}_M = \sum_{i=0}^4 \Omega^{i-4} \beta_i {\cal E}_i~,
\end{align}
where $\beta_i = (-1)^i [(4-i)(3-i)/2 + (4-i)\alpha_3 + \alpha_4 ]$. In this
expression, based on the transformed matrix $\mathbb{\tilde X} \equiv
\sqrt{\tilde{g}^{-1}f}$, we have introduced the elementary symmetric
polynomial ${\cal E}_i$ as:
\begin{eqnarray}\label{massstructure}
 {\cal E}_0 = 1 ~,~
 {\cal E}_1 = \mathbb{\tilde X}^\mu_{\mu} ~,~
 {\cal E}_2 = \mathbb{\tilde X}^\mu_{[\mu} \mathbb{\tilde X}^\nu_{\nu]} ~,~
\nonumber\\
 {\cal E}_3 = \mathbb{\tilde X}^\mu_{[\mu} \mathbb{\tilde X}^\nu_{\nu}
\mathbb{\tilde X}^\sigma_{\sigma]} ~,~
 {\cal E}_4 = \mathbb{\tilde X}^\mu_{[\mu} \mathbb{\tilde X}^\nu_{\nu}
\mathbb{\tilde X}^\sigma_{\sigma} \mathbb{\tilde X}^\rho_{\rho]} ~.
\end{eqnarray}
Then the resulting Lagrangian in the Einstein frame can be written as
\begin{eqnarray}
\label{Einsteinframe}
 {\cal L} = \sqrt{|\tilde{g}|} \bigg[ \frac{M_p^2}{2} ( \tilde{R} + 2m_g^2
\tilde{\cal U}_M ) - \frac{1}{2}\partial_\mu\varphi\partial^\mu\varphi -
U(\varphi) \bigg]~.
\end{eqnarray}

At first sight one might feel that our construction has a relation with the
quasi-dilaton massive gravity \cite{D'Amico:2012zv} and the mass-varying
massive gravity \cite{Huang:2012pe,Saridakis:2012jy}. However, these
scenarios are radically different, straightaway from the starting point of
the model-building, and moreover they obey completely different symmetries.
In particular, in the quasi-dilaton massive gravity the coefficient in front
of the kinetic term of the scalar field is a free parameter, while in our
model it results to be unity, and this feature has a crucial effect on the
perturbational analysis, reducing the number of degrees of freedom as we will
see. Additionally, while in mass-varying massive gravity the separate
gravitational terms acquire a common overall factor, in the present
construction they result obtaining different scalar-field dependencies, which
make the two models radically different.

\section{Hamiltonian constraint analysis}

To examine the BD ghost issue, one must perform the Hamiltonian constraint
analysis \cite{Hassan:2011ea}. For simplicity we work within the Einstein
frame and expand the metrics using the famous Arnowitt-Deser-Misner (ADM)
formalism:
\begin{small}
\begin{eqnarray}
 \tilde{g}_{\mu\nu} dx^\mu dx^\nu = -(N_g^0)^2dt^2+\gamma_{ij} (dx^i+N_g^idt)
(dx^j+N_g^jdt),\nonumber \\
  f_{\sigma\rho} dx^\sigma dx^\rho = -(N_f^0)^2dt^2+\omega_{ab}
(dx^a+N_f^adt) (dx^b+N_f^bdt).
\end{eqnarray}
\end{small}
The lapse $N_g^0$ and shift $\vec{N}_g$ (the three $N_g^i$'s expressed as
vector) of the physical metric, as well as the corresponding ones for the
fiducial metric, $N_f^0$ and $\vec{N}_f$ respectively, are all non-dynamical.
In massive gravity $\gamma_{ij}$ allows for at most six propagating modes,
one of them being the origin of the BD ghost. A potentially healthy theory
must maintain a single constraint on ${\bar{\gamma}}$ (from now on a bar
denotes the matrix form) and the conjugate momenta, along with an associated
secondary constraint, which will lead to the elimination of the ghost DoF. In
the following we briefly show the existence of these constraints in $F(R)$
massive gravity, one can find additional details in \cite{Cai:2014upa}.

In order to introduce the Lagrange multiplier explicitly, we define a new
shift $\vec{\eta}$ through $\vec{N}_g-\vec{N}_f = ( N_f^0 \mathbb{I} +N_g^0
\mathbb{D} )\vec{\eta}$, where $\mathbb{I}$ is the $3\times3$ unit matrix and
the $3\times3$ matrix $\mathbb{D}$ satisfies $\lambda \mathbb{D}\mathbb{D} =
[\bar{\gamma}^{-1} -(\mathbb{D} \vec{\eta}) (\mathbb{D} \vec{\eta})^{\rm T} ]
\bar\omega$ with $\lambda = 1-{\vec{\eta}}^{\rm T}\bar\omega{\vec{\eta}}$.
The conjugate momenta are defined as $\pi \equiv \frac{\delta S_{\rm
G}}{\delta \dot\varphi}$ and $\bar{\Pi} \equiv\frac{\delta S_{\rm G}}{\delta
\dot{\bar\gamma}}$. Then we can derive the Hamiltonian as:
 \begin{eqnarray}
  {\rm H} = \int d^3x [ {\cal H} - N_g^0 \mathcal{C}(\varphi, \pi,
\bar{\gamma}, \bar{\Pi}, \vec{\eta}) ],
 \end{eqnarray}
where
\begin{align}
 &{\cal H} =
-(N_f^0\vec{\eta}+\vec{N}_f)\vec{\mathcal{R}}-N_f^0\sqrt{|\bar\gamma|}
\mathfrak{F}_{\rm H} m_g^2M_p^2~,\\
 &\mathcal{C} = \mathcal{R} +\vec{\mathcal{R}}^{\rm T} \mathbb{D} \vec{\eta}
+ \sqrt{|\bar\gamma|} \mathfrak{F}_{\rm C} m_g^2M_p^2~.
\end{align}
Here we have introduced the coefficients:
 \begin{eqnarray}
&& \mathcal{R}=\frac{\sqrt{|\bar\gamma|}}{2} \left[ M_p^2 R_3
-\varphi_{,i}\varphi^{,i} -U(\varphi) \right]\nonumber\\
&&\ \ \ \ \ \ \ +
\frac{1}{ {\sqrt{|\bar\gamma|}}}
\left[ \frac{ {({\rm
Tr}\bar{\Pi})}^2 }{M_p^2} -\frac{ 2{\bar{\Pi}}^2 }{M_p^2} - \frac{\pi^2}{2}
\right]
 \end{eqnarray}
 and
 \begin{eqnarray}
 {\mathcal{R}}_i = 2 \gamma_{ij}\Pi^{jk}_{;k} -
\pi\varphi_{,i},
 \end{eqnarray}
while the coefficients appearing in the mass terms are
\begin{small}
\begin{align}
 \mathfrak{F}_{\rm H} &= \frac{\beta_1\lambda^{\frac{1}{2}}}{\Omega^{3}} +
\frac{\beta_2}{\Omega^{2}} \left[\lambda{\rm Tr}\mathbb{D} +\vec{\eta}^{\rm
T}\bar\omega\mathbb{D}\vec{\eta}\right]
 \nonumber\\
 &\ \ \, + \frac{\beta_3}{\Omega} \left[2\lambda^{\frac{1}{2}}
\mathbb{D}^{[l}_l\eta^{i]}\omega_{ij}\mathbb{D}^j_k\eta^k +
\lambda^{\frac{3}{2}} \mathbb{D}^{[i}_i\mathbb{D}^{j]}_j\right] + \frac{
\beta_4 |\bar\omega|^{\frac{1}{2}} }{ |\bar\gamma|^{\frac{1}{2}} } ~,\\
 \mathfrak{F}_{\rm C} &= \frac{\beta_0}{\Omega^{4}} +
\frac{\beta_1\lambda^{\frac{1}{2}}{\rm Tr}\mathbb{D}}{\Omega^{3}} +
\frac{\beta_2\lambda \mathbb{D}^{[i}_i\mathbb{D}^{j]}_j }{\Omega^{2}} +
\frac{\beta_3\lambda^{\frac{3}{2}}}{\Omega}
\mathbb{D}^{[i}_i\mathbb{D}^{j}_j\mathbb{D}^{k]}_k ~.
\end{align}
\end{small}

Varying the Hamiltonian with respect to the new shift $\vec{\eta}$ does not
yield any constraints, but due to this new shift vector the variation with
the lapse function $N_g^0$ does give a constraint which reads,
\begin{eqnarray}
 \mathcal{C}(\varphi, \pi, \bar{\gamma}, \bar{\Pi}, \vec{\eta}) = 0~.
\end{eqnarray}
In order for this constraint to hold at all times we must also demand that
\begin{equation}\label{c2const}
\mathcal{C}^{(2)}=\{ \mathcal{C} , {\rm H}\}=0,
\end{equation}
where the Poisson brackets of two quantities are defined as
\begin{align}
\{\mathcal{O}_1(x),\mathcal{O}_2(y)\}=\sum_i\int d^3z
\Big[&\frac{\delta\mathcal{O}_1(x)}{\delta
q_i}\frac{\delta\mathcal{O}_2(y)}{\delta p^i}\nonumber\\
&-\frac{\delta\mathcal{O}_1(x)}{\delta
p^i}\frac{\delta\mathcal{O}_2(y)}{\delta q_i}\Big],
\end{align}
with the $q_i$ being the canonical variables $(\gamma_{ij},\varphi)$ and
$p^i$ their conjugate momenta $(\Pi^{ij},\pi)$.

 Equation (\ref{c2const}) must generate a needed second constraint on
$\varphi$, $\bar{\gamma}$ and their conjugate momenta, therefore it must not
vanish identically and/or must not be an equation that determines the lapse
function $N_g^0$. The latter condition will not hold if  $\{ \mathcal{C}(x),
\mathcal{C}(y)\}$ does not vanish as it appears in $\mathcal{C}^{(2)}$ with
$N_g^0$ as its coefficient. Fortunately, one can show that $\{
\mathcal{C}(x), \mathcal{C}(y)\}$=0 identically
\cite{Hassan:2011ea,Cai:2014upa}.
The remaining term $
\mathcal{C}^{(2)}=\{ \mathcal{C}, \int d^3x {\cal H}(x) \}$ becomes
 \begin{align}
\mathcal{C}^{(2)}=&
\mathcal{C}\nabla_i(N_f\eta^i+N_f^i)+m_g^2M_p^2(\gamma_{mn}\Pi^k_k-2\Pi_{mn}
)\mathfrak{F}_{\rm H}^{mn} \nonumber \\
 &+m_g^2M_p^2 N_f\mathbb{D}^i_k \eta^k\Big(2\sqrt{\gamma}(\nabla_m F_{\rm
H}^{mn})\gamma_{ni}-\frac{\partial \mathfrak{F}_{\rm
H}}{\partial\varphi}\partial_i\varphi\Big)\nonumber\\
 &+\nabla_i(N_f\eta^i+N_f^i)(\mathcal{R}_j\mathbb{D}
^j_k\eta^k-m_g^2M_p^2\sqrt{\gamma}\gamma_{jk}B^{kj})\nonumber\\
 &-m_g^2M_p^2(N_f\eta^i+N_f^i)\sqrt{\gamma}\frac{\partial
\mathfrak{F}_{\rm C}}{\partial\varphi}\partial_i\varphi,
 \end{align}
 with $F_{\rm
H}^{mn}=\frac{1}{\sqrt{\gamma}}\frac{\partial(\sqrt{\gamma}\mathfrak{F}_{\rm
H})}{\partial\gamma_{mn}}$ and
\begin{align}
 B^{ki}=&\gamma^{km}\Big[\frac{\beta_1}{
\lambda^{1/2}}\omega_{ma}(\mathbb{D}^{-1})^a_j+\beta_2\big(\omega_{ma}
(\mathbb{D}^{-1})^a_j\mathbb{D}^b_b-\omega_{mj}\big)\nonumber \\
 &+\beta_3\lambda^{1/2}\big[\omega_{ma}(\mathbb{D}^{-1})^a_j-\omega_{mj}
(\mathbb{D}^{-1})^a_a\big]\nonumber\\
 &+\frac{\beta_3\lambda^{1/2}}{2}\omega_{ma}(\mathbb{D}^{-1})^a_j
(\mathbb{D}^a_a\mathbb{D}^b_b-\mathbb{D}^a_b\mathbb{D}^b_a)\Big]\gamma^{ji}.
\end{align}
From this expression we see that $\mathcal{C}^{(2)}$ contains no mention
of $N_g^0$ and is not proportional to the original constraint $\mathcal{C}$,
hence it does not vanish identically when $\mathcal{C}=0$. Therefore
imposing
\begin{eqnarray}
 \{ \mathcal{C}, \int d^3x {\cal H}(x) \}=0~,
\end{eqnarray}
gives a second nontrivial constraint. This result is not too surprising as the effect of considering $F(R)$ gravity amounts to (in the Einstein frame) multiplying each of the graviton's mass terms by a power of the conformal factor $\Omega$. Therefore the structure of Eq. (\ref{UMs}) which is responsible for the elimination of the BD ghost is still preserved as seen in Eq. (\ref{massstructure}).

Now since $\mathcal{C}^{(2)}=0$ must remain valid at all times one must make sure that the equation $\{\mathcal{C}^{(2)},H\}=0$ does not lead to additional constraints but instead give an equation determining $N_g^0$. This is the case only if $\{\mathcal{C}^{(2)}, \int d^3x {\cal H}(x)\}\ne 0$ and $\{\mathcal{C}^{(2)}, \mathcal{C}\}\ne 0$. These two conditions are satisfied by the Fierz-Pauli constraints for which $ \mathcal{C}^{(2)}$ and $\mathcal{C}$ reduces to at lowest order in the $\gamma_{ij},~\pi_{ij}$ fields and with $\varphi=0$. Therefore as argued in \cite{Hassan:2011ea},
considering the terms of higher order in the fields cannot change $\{\mathcal{C}^{(2)}, \int d^3x {\cal H}(x)\}$ and $\{\mathcal{C}^{(2)}, \mathcal{C}\}$ in a way that makes them vanish identically. Additionally, since $\varphi=0$ belongs to the constraining surface, adding this new DoF will not change the fact that no tertiary constraint exists. Hence we see that the effect of considering an $F(R)$ modification does not lead to a resurgence of the BD ghost.

\section{Cosmology}

When applied in cosmological frameworks, the scenario of $F(R)$ massive
gravity exhibits a large class of phenomenological behaviors due to the
combination of the $F(R)$ and graviton-mass sectors. Let us start with a
Minkowski fiducial metric $f_{\sigma\rho}=\eta_{\sigma\rho}$. The model
allows only for open Friedmann-Robertson-Walker (FRW) universe, and thus we
consider the physical metric in Jordan frame as
\begin{eqnarray}
 ds^2=-N^2dt^2+a^2(t) \gamma^{K}_{ij}dx^idx^j ~,
\end{eqnarray}
with $\gamma^{K}_{ij} dx^idx^j = \delta_{ij}dx^idx^j -
\frac{a_0^2(\delta_{ij}x^idx^j)^2}{1-a_0^2\delta_{ij}x^ix^j}$ and
$a_0=\sqrt{|K|}$ is associated with the spatial curvature. The St\"uckelberg
scalars are:  $\varphi^0 = b(t) \sqrt{ 1 + a_0^2 \delta_{ij}x^ix^j }$,
$\varphi^i = a_0 b(t) x^i$. Then the polynomials defined in \eqref{UMs} take
the forms of
\begin{align}
 {\cal U}_2 &=3a(a-a_0 b)(2{N}{a}-{b'}{a}-{N}a_0 b)~,\nonumber\\
 {\cal U}_3 &=({a}-a_0 b)^2(4{N}{a}-3{a}{b'}-Na_0  b)~,\nonumber\\
 {\cal U}_4 &=({a}-a_0 b)^3({N}-{b'})~,
\end{align}
where primes denote derivatives with respect to $t$. Finally, for simplicity
we assume that the gravitational sector couples minimally to the regular
matter component.

Variation of the action with respect to $b$, $N$ and $a$ gives respectively
the constraint and the two Friedmann equations, namely
\begin{align}
\label{constraint0}
 &(\dot{a}-a_0)Y_1 = 0 ~, \\
 &3 M_p^{2} F_{,R} \left( H^2-\frac{a_0^2}{a^2} \right) = \rho_m+\rho_{\rm
IR}+\rho_{\rm UV} ~,\\
 &M_p^{2} F_{,R}\left(-2\dot{H}-3H^2 +\frac{a_0^2}{a^2}\right) = p_m+p_{\rm
IR}+p_{\rm UV} ~,
\end{align}
with $\dot{a}=\frac{a'}{N}$ and $H=\frac{\dot{a}}{a}$. In the above
expressions we have defined the IR (massive gravity) effective contribution
\begin{align}
 & \rho_{\rm IR} = m_g^2M_p^2({\cal B}-1)(Y_1+Y_2)~, \nonumber\\
 & p_{\rm IR} = -m_g^2M_p^2({\cal B}-1)Y_2-m_g^2M_p^2(\dot{b}-1)Y_1~,
\end{align}
as well as the UV ($F(R)$ sector) effective contribution
{\small{
\begin{align}
\label{rhofR}
 &\rho_{\rm UV} = M_p^2 \left[\frac{R F_{,R}-F}{2}-3H\dot{R} F_{,RR}\right]~,
\\
\label{PfR}
 & p_{\rm UV} = M_p^2 \left[\dot{R}^2 F_{,RRR}+2H\dot{R} F_{,RR}+\ddot{R}
F_{,RR} +\frac{F-RF_{,R}}{2} \right]~,
\end{align}}}
where the polynomials $Y_{1,2}$ are given by
$Y_1 = (3-2{\cal B})+\alpha_3(3-{\cal B})(1-{\cal B})+\alpha_4(1-{\cal B})^2$
and
$Y_2 = (3-{\cal B})+\alpha_3(1-{\cal B})$, with ${\cal B}=\frac{a_0b}{a}$.

Similar to all massive gravity scenarios, Eq. \eqref{constraint0} constrains
the dynamics significantly. As in self-accelerating backgrounds of dRGT
\cite{Gumrukcuoglu:2011ew}, the nontrivial solutions correspond to the case
of $Y_1=0$ and yield
\begin{align}
 {\cal{B}}_\pm = \frac{1 +2\alpha_3 +\alpha_4\pm\sqrt{1 +\alpha_3 +\alpha_3^2
-\alpha_4}}{\alpha_3 +\alpha_4}~.
\end{align}
This relation can be always fulfilled by choosing $b(t)\propto a(t)$, and
therefore it yields $\rho_{\rm IR}=-p_{\rm IR}$ to be constant, as it is
expected similarly to standard nonlinear massive gravity
\cite{Koyama:2011xz}. However, the crucial issue is that in the present model
the remaining $F(R)$ sector can be taken at will, leading to a large class of
cosmologies. Amongst them, an interesting class is when the $F(R)$ sector is
important at early times and thus responsible for inflation, while the
massive graviton is dominant at late times and can drive the universe
acceleration as observed today.

\begin{figure}[ht]
\begin{center}\!\!\!\!\!\!\!\!\!\!
\mbox{\epsfig{figure=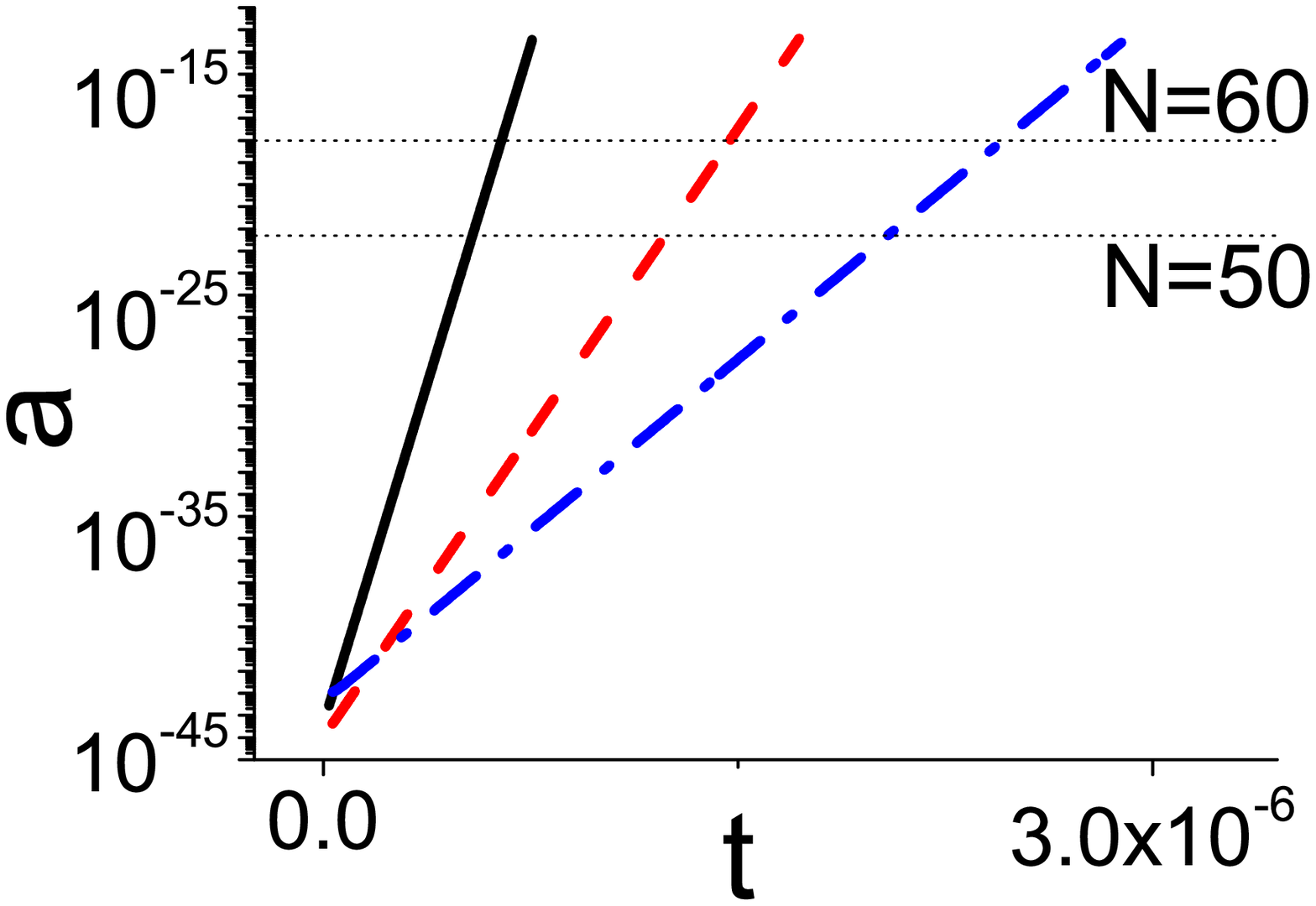,width=4.54cm,angle=0}}
\!\!\!\!\!\!\!\!\!\!\!\! \!\!
\mbox{\epsfig{figure=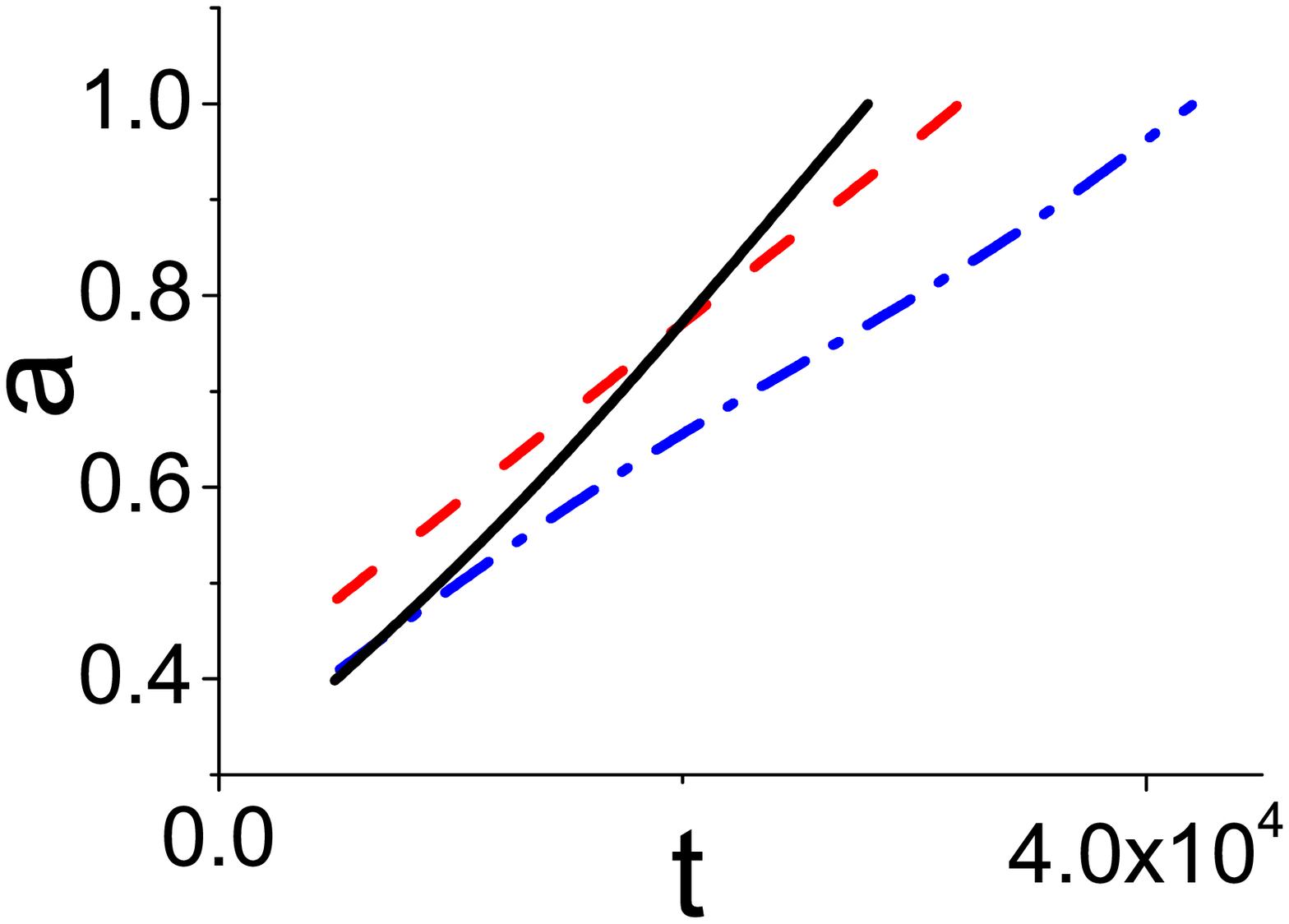,width=4.54cm,angle=0}}
\caption{ The left panel presents three inflationary solutions corresponding
to
a) $m_g=10^{-50}$, $\alpha_3=2$, $\alpha_4=-1$, $a_0=5\times 10^{-41}$,
$\xi=10^{10}$ (black-solid),
b) $m_g=10^{-50}$, $\alpha_3=10$, $\alpha_4=10$, $a_0=10^{-41}$, $\xi=10^{9}$
(red-dashed),
c) $m_g=10^{-50}$, $\alpha_3=1$, $\alpha_4=1$, $a_0= 10^{-40}$, $\xi=10^{10}$
(blue-dash-doted).
 The two horizontal lines mark the $N=50$ and $N=60$ e-folding regimes. All
parameters are in Planck units. The right panel depicts three late-time
accelerating solutions corresponding to
a) $m_g=3$, $\alpha_3=3$, $\alpha_4=-5$, $a_0=0.05$, $\xi=0.5$ (black-solid),
b) $m_g=1.5$, $\alpha_3=1$, $\alpha_4=-2$, $a_0=0.01$, $\xi=0.5$
(red-dashed),
c) $m_g=3$, $\alpha_3=10$, $\alpha_4=1$, $a_0=0.05$, $\xi=0.5$
(blue-dash-doted).
 All parameters are in units where the present Hubble parameter is $H_0=1$,
and we have imposed $\Omega_{m0}\approx0.31$, $\Omega_{DE0}\approx0.69$,
$\Omega_{k0}\approx0.01$ at the present scale factor $a_0=1$.
} \label{Fig1}
\end{center}
\end{figure}

In order to provide a representative example we consider the well-known
Starobinsky model with $F(R)=R+ \frac{\xi}{M_p^2} R^2$ in numerical
estimates. In the left panel of Fig.~\ref{Fig1} we present the early-time
inflationary solutions for three parameter choices, while in the right panel
we depict the late-time self-accelerating solutions. In this particular
choice, the Ricci scalar becomes very small at late times and thus the
$F(R)$'s contribution is dramatically suppressed by the Planck scale.
Therefore, only the massive-gravity part contributes to the late-time
acceleration. However, note that in the general case the total effective dark
energy constitutes of both the massive gravity  as well as the
$F(R)$-modification sectors, that is $\rho_{\rm DE}\equiv\rho_{\rm
IR}+\rho_{\rm UV}$. Therefore, our model is expected to be very interesting
phenomenologically \cite{Cai:2014upa}.

\section{Perturbation analysis}

The scenario at hand is free of the BD ghost and its cosmological
applications allow for a large class of behaviors. However, the last and
necessary step is to examine whether such cosmological applications remain
free of instabilities at the perturbative level, which is exactly the weak
and disastrous point of standard nonlinear massive gravity pointed out in
\cite{DeFelice:2012mx} (see also \cite{Gumrukcuoglu:2011zh,
Andrews:2013ora}). In the rest of the Letter we briefly show that the scalar
perturbations in our model can be stable at the linear perturbative level
under certain parameter choices.

For simplicity we work in the Einstein frame, using the Lagrangian
(\ref{Einsteinframe}), and then consider perturbations around a homogeneous
and isotropic background. The scalar perturbations of our variables involve
the metric part
\begin{align}
 & \delta g_{00} = -2\tilde{N}^2\phi~,~~\delta g_{0i} = \tilde{N} \tilde{a}
\partial_i B ~,~
 \nonumber\\
 & \delta g_{ij} = \tilde{a}^2 [2\gamma^{K}_{ij}\psi
+(\nabla_i\nabla_j-\frac{1}{3}\gamma^K_{ij}\nabla_k\nabla^k)]E ~,
\end{align}
and the field fluctuation $\delta\varphi$. Using the Hamiltonian and momentum
constraints, as well as the background equations of motion, we can integrate
out the non-dynamical modes, namely $\phi$, $B$ and $E$. Therefore, the
would-be BD ghost is eliminated in our model. Furthermore, since the scalar
DoF of the graviton is non-dynamical at the linear level on the
self-accelerating solution, one can introduce the usual Bardeen potential
$\psi_B$ and define a generalized Mukhanov-Sasaki variable
\begin{eqnarray}
Q \equiv
\delta\varphi_B + \frac{\dot\varphi\psi_B}{H}.
\end{eqnarray}
This allows us to obtain the
perturbation equation of our single propagating scalar DoF in the Fourier
space as
\begin{eqnarray}\label{EoMQ}
 \ddot Q_k +3 H\dot Q_k + \left[\frac{k^2}{a^2} +U_{,\varphi\varphi} -
\frac{1}{M_p^2a^3}(\frac{a^3}{H} \dot\varphi^2)^{\cdot}\right] Q_k
 \nonumber\\
 \ \ = \frac{2m_g^2\tilde{Y}_Q}{3\Omega^4} Q_k - \frac{2K}{a^2H^2}
\left(\ddot\varphi -\frac{\dot{H} \dot\varphi}{H} \right) \psi_B ~,
\end{eqnarray}
where $\tilde{Y}_Q\equiv 4(1-\tilde{\cal B}) \tilde{Y}_2$ is defined in the
Einstein frame. Note that $Q$ is the only dynamical perturbation variable
since $\psi_B$ can be determined by it as well.

From the above analysis one can clearly see the qualitative difference of the
present construction, comparing to other extended nonlinear massive gravity
models, such are the quasi-dilaton massive gravity \cite{D'Amico:2012zv} and
the mass-varying massive gravity \cite{Huang:2012pe}. In particular, these
extensions involve two extra scalar DoF, as it can be verified by counting
the number of nonzero eigenvalues of the matrix for the kinetic part of the
perturbation action \cite{Gumrukcuoglu:2013nza}. Applying the method of
\cite{Gumrukcuoglu:2013nza} in the present scenario, by setting the
coefficient in front of the scalar-field kinetic term to unity, we find that
there exists only one nonzero eigenvalue, and this implies only a single DoF.
A detailed analysis of this issue can be found in the accompanied paper
\cite{Cai:2014upa}.

The l.h.s. of the perturbation equation \eqref{EoMQ} is exactly the same as
the usual one in GR plus a scalar field, but the r.h.s. involves a mass term
due to the graviton potential. Its positivity depends on the coefficient
$\tilde{Y}_Q$ and directly determines whether the model suffers from a
tachyonic instability or not. Obviously, a healthy model of $F(R)$ massive
gravity requires $\tilde{Y}_Q<0$, which provides the corresponding allowed
regime of the parameter space. Additionally, the last term of \eqref{EoMQ}
appears due to the spatial curvature. Since this term would easily dilute out
along the cosmic expansion, it is harmless to the model when applied into
cosmology. Therefore, we conclude that there exists enough parameter space
for scalar perturbations to be stable throughout the cosmological evolutions
of phenomenological interest.

\section{Conclusions}

The study of massive gravity may be important in understanding the observed
acceleration of present cosmic expansion, which is one of the greatest
mysteries in modern physics. In this regard, the question on establishing a
theoretically healthy and observationally viable model of nonlinear massive
gravity has attracted the interest of the literature.

The theory of $F(R)$ nonlinear massive gravity, as a possible GR modification
both at the IR and UV regimes, has significant advantages both at the
theoretical as well as at the cosmological levels. Firstly, due to the usual
dRGT-like graviton potential it inherits its benefits and is free of BD
ghosts. Furthermore, due to the freedom of the $F(R)$ sector combined with
the graviton-mass, it allows for a large class of cosmological evolutions.
For instance a simple $R^2$ form is able to drive both early universe
inflation and late-time acceleration, determining the whole cosmic evolution
in a unified way.

We would like to end by highlighting the advantage of our model that the
perturbations around a cosmological background can be stabilized due to the
$F(R)$ term, which introduces a scalar DoF at the linear level, and hence it
constrains the scalar metric perturbations to be as in GR. Usually, the
nonlinear inclusion of the gravitational mass gives rise to a scalar DoF,
that is the longitudinal graviton. Although the inclusion of the $F(R)$
sector at first introduces another scalar mode, this mode nicely ``eats'' the
nonlinear mode due to the graviton mass, and moreover it imposes the
constraint on the stability. This mechanism is very similar to the process of
the spontaneous symmetry breaking of particle physics governed by the {\it
Goldstone theorem}. In this respect, the possible instabilities that could
appear at higher nonlinear regime, do not appear unless the perturbation
theory itself break down. Additionally, we mention that once the
perturbations evolve into nonlinear regime, higher curvature terms
would become important accompanied by the high energy scale, and thus
completely change the dynamics of the theory. The above features may reveal
that the UV and IR behaviors of gravitation may not be independent.

The possibility of a gravitational Goldstone theorem deserves further
investigation. In particular, since the $F(R)$ sector can be reformulated as
a scalar field minimally coupled to the Ricci scalar with an effective
potential, and since for a wide class of $F(R)$ forms the effective potential
approaches an extremely flat plateau in the UV regime, then from the
viewpoint of symmetry the corresponding effective potential indicates an
approximately shift symmetry along the scalar field. When the scalar field
evolves into the IR regime it is stabilized at the vacuum and therefore the
shift symmetry can be spontaneously broken. One may make an analogue with the
scalar field and the dilaton. Thus, one at first expects the second
propagating scalar mode to appear in the IR regime, however, it was eaten by
the dilaton field through the process of the spontaneous shift symmetry
breaking. This interesting property is perhaps an indication for the
aforementioned possibility of a gravitational Goldstone theorem.

\begin{acknowledgments}
We are grateful to R. Brandenberger, P. Chen, A. De Felice, S. Deser,
G. Gabadadze, A. E. G\"umr\"uk\c{c}\"uo\u{g}lu, C. Lin, S. Mukohyama,
and M. Trodden, for useful discussions. The work of YFC and FD is
supported in part by an NSERC Discovery grant and by funds from the Canada
Research Chair program. FD was also supported by an FQRNT B2 scholarship
during the final stage of this work. The research of E.N.S. is implemented
within the framework of the Operational Program ``Education and Lifelong
Learning'' (Actions Beneficiary: General Secretariat for Research and
Technology), and is co-financed by the European Social Fund (ESF) and the
Greek State.
\end{acknowledgments}


\begin{thebibliography}{99}

\bibitem{Fierz:1939ix}
  M.~Fierz, W.~Pauli,
  Proc.\ Roy.\ Soc.\ Lond.\ A {\bf 173}, 211 (1939).

\bibitem{vanDam:1970vg}
  H.~van Dam and M.~J.~G.~Veltman,
  Nucl.\ Phys.\ B {\bf 22}, 397 (1970);
  V.~I.~Zakharov,
  JETP Lett.\  {\bf 12}, 312 (1970).

\bibitem{Vainshtein:1972sx}
  A.~I.~Vainshtein,
  Phys.\ Lett.\  B {\bf 39}, 393 (1972).

\bibitem{Boulware:1973my}
  D.~G.~Boulware, S.~Deser,
  Phys.\ Rev.\ D {\bf 6}, 3368 (1972).

\bibitem{ArkaniHamed:2002sp}
  N.~Arkani-Hamed, H.~Georgi and M.~D.~Schwartz,
  Annals Phys.\  {\bf 305}, 96 (2003).

\bibitem{deRham:2010ik}
  C.~de Rham and G.~Gabadadze,
  Phys.\ Rev.\ D {\bf 82}, 044020 (2010);
  C.~de Rham, G.~Gabadadze and A.~J.~Tolley,
  Phys.\ Rev.\ Lett.\  {\bf 106}, 231101 (2011).

\bibitem{Hassan:2011ea}
  S.~F.~Hassan and R.~A.~Rosen,
  JHEP {\bf 1204}, 123 (2012);
  S.~F.~Hassan and R.~A.~Rosen,
  Phys.\ Rev.\ Lett.\  {\bf 108}, 041101 (2012).


\bibitem{deRham:2010tw}
  C.~de Rham, G.~Gabadadze, L.~Heisenberg and D.~Pirtskhalava,
  Phys.\ Rev.\ D {\bf 83}, 103516 (2011);
  G.~D'Amico, C.~de Rham, S.~Dubovsky, G.~Gabadadze, D.~Pirtskhalava and
A.~J.~Tolley,
  Phys.\ Rev.\ D {\bf 84}, 124046 (2011);
  K.~Koyama, G.~Niz and G.~Tasinato,
  JHEP {\bf 1112}, 065 (2011).


\bibitem{Gumrukcuoglu:2011ew}
  A.~E.~Gumrukcuoglu, C.~Lin and S.~Mukohyama,
  JCAP {\bf 1111}, 030 (2011);
  A.~E.~Gumrukcuoglu, C.~Lin and S.~Mukohyama,
  Phys.\ Lett.\ B {\bf 717}, 295 (2012);
  A.~De Felice, A.~E.~Gumrukcuoglu and S.~Mukohyama,
  Phys.\ Rev.\ D {\bf 88}, 124006 (2013).


 \bibitem{Comelli:2011zm}
  D.~Comelli, M.~Crisostomi, F.~Nesti and L.~Pilo,
  JHEP {\bf 1203}, 067 (2012)
  [Erratum-ibid.\  {\bf 1206}, 020 (2012)];
  D.~Comelli, M.~Crisostomi and L.~Pilo,
  JHEP {\bf 1206}, 085 (2012);
  V.~F.~Cardone, N.~Radicella and L.~Parisi,
  Phys.\ Rev.\ D {\bf 85}, 124005 (2012);
  P.~Gratia, W.~Hu and M.~Wyman,
  Phys.\ Rev.\ D {\bf 86}, 061504 (2012);
  T.~Kobayashi, M.~Siino, M.~Yamaguchi and D.~Yoshida,
  Phys.\ Rev.\ D {\bf 86}, 061505 (2012);
  D.~Langlois and A.~Naruko,
  Class.\ Quant.\ Grav.\  {\bf 29}, 202001 (2012);
  Y.~-F.~Cai, D.~A.~Easson, C.~Gao and E.~N.~Saridakis,
  Phys.\ Rev.\ D {\bf 87}, 064001 (2013);
  Y.~-l.~Zhang, R.~Saito and M.~Sasaki,
  JCAP {\bf 1302}, 029 (2013);
  G.~Leon, J.~Saavedra and E.~N.~Saridakis,
 Class.\ Quant.\ Grav.\  {\bf 30}, 135001 (2013);
  K.~Hinterbichler, J.~Stokes and M.~Trodden,
  Phys.\  Lett.\ B {\bf 725}, , 1 (2013);
  K.~Zhang, P.~Wu and H.~Yu,
 Phys.\ Rev.\ D {\bf 87}, 063513 (2013);
  M.~S.~Volkov,
 Class.\ Quant.\ Grav.\  {\bf 30}, 184009 (2013);
  G.~Tasinato, K.~Koyama and G.~Niz,
 Class.\ Quant.\ Grav.\  {\bf 30}, 184002 (2013);
  H.~Li and Y.~Zhang,
 arXiv:1304.4780 [gr-qc];
  N.~Khosravi, G.~Niz, K.~Koyama and G.~Tasinato,
 JCAP {\bf 1308}, 044 (2013);
  T.~Q.~Do and W.~F.~Kao,
  Phys.\ Rev.\ D {\bf 88}, 063006 (2013);
  M.~Sasaki, D.~-h.~Yeom and Y.~-l.~Zhang,
  Class.\ Quant.\ Grav.\  {\bf 30}, 232001 (2013);
  Y.~-l.~Zhang, R.~Saito, D.~-h.~Yeom and M.~Sasaki,
  arXiv:1312.0709 [hep-th];
  S.~I.~Vacaru,
  arXiv:1401.2882 [physics.gen-ph].


\bibitem{DeFelice:2012mx}
  A.~De Felice, A.~E.~Gumrukcuoglu and S.~Mukohyama,
  Phys.\ Rev.\ Lett.\  {\bf 109}, 171101 (2012);
  A.~De Felice, A.~E.~Gumrukcuoglu, C.~Lin and S.~Mukohyama,
  JCAP {\bf 1305}, 035 (2013);
  A.~De Felice, A.~E.~Gumrukcuoglu, C.~Lin and S.~Mukohyama,
  Class.\ Quant.\ Grav.\  {\bf 30}, 184004 (2013).




\bibitem{Utiyama:1962sn}
  R.~Utiyama and B.~S.~DeWitt,
  J.\ Math.\ Phys.\  {\bf 3}, 608 (1962).

\bibitem{Stelle:1976gc}
  K.~S.~Stelle,
  Phys.\ Rev.\ D {\bf 16}, 953 (1977).
%

\bibitem{Deser:1977nt}
  S.~Deser, J.~H.~Kay and K.~S.~Stelle,
  Phys.\ Rev.\ Lett.\  {\bf 38}, 527 (1977).

\bibitem{Vilkovisky:1992pb}
  G.~A.~Vilkovisky,
  Class.\ Quant.\ Grav.\  {\bf 9}, 895 (1992).

\bibitem{DeFelice:2010aj}
  A.~De Felice and S.~Tsujikawa,
  Living Rev.\ Rel.\  {\bf 13}, 3 (2010).

\bibitem{Starobinsky:1980te}
  A.~A.~Starobinsky,
  Phys.\ Lett.\ B {\bf 91}, 99 (1980).

\bibitem{Ade:2013uln}
  P.~A.~R.~Ade {\it et al.},
  arXiv:1303.5082 [astro-ph.CO].

\bibitem{Nojiri:2012re}
  S.~'i.~Nojiri, S.~D.~Odintsov and N.~Shirai,
  JCAP {\bf 1305}, 020 (2013).


\bibitem{D'Amico:2012zv}
  G.~D'Amico, G.~Gabadadze, L.~Hui and D.~Pirtskhalava,
  Phys.\ Rev.\ D {\bf 87}, 064037 (2013);
   R.~Gannouji, M.~. W.~Hossain, M.~Sami and E.~N.~Saridakis,
 Phys.\ Rev.\ D {\bf 87}, 123536 (2013).


\bibitem{Huang:2012pe}
  Q.~-G.~Huang, Y.~-S.~Piao and S.~-Y.~Zhou,
  Phys.\ Rev.\ D {\bf 86}, 124014 (2012).

\bibitem{Saridakis:2012jy}
  E.~N.~Saridakis,
  Class.\ Quant.\ Grav.\  {\bf 30}, 075003 (2013);
  Y.~-F.~Cai, C.~Gao and E.~N.~Saridakis,
  JCAP {\bf 1210}, 048 (2012);
  D.~-J.~Wu, Y.~-S.~Piao and Y.~-F.~Cai,
  Phys.\ Lett.\ B {\bf 721}, 7 (2013).

\bibitem{Cai:2014upa}
  Y.~-F.~Cai and E.~N.~Saridakis,
    Phys.\ Rev.\ D {\bf 90}, 063528 (2014).




\bibitem{Andrews:2013ora}
  M.~Andrews, G.~Goon, K.~Hinterbichler, J.~Stokes and M.~Trodden,
  Phys.\ Rev.\ Lett.\  {\bf 111}, 061107 (2013);
  Q.~-G.~Huang, K.~-C.~Zhang and S.~-Y.~Zhou,
  JCAP {\bf 1308}, 050 (2013);
%
  A.~De Felice and S.~Mukohyama,
  Phys.\ Lett.\ B {\bf 728}, 622 (2014);
%
  M.~Andrews, K.~Hinterbichler, J.~Stokes and M.~Trodden,
Galileons,''
  Class.\ Quant.\ Grav.\  {\bf 30}, 184006 (2013).


\bibitem{Koyama:2011xz}
  K.~Koyama, G.~Niz and G.~Tasinato,
  Phys.\ Rev.\ Lett.\  {\bf 107}, 131101 (2011).

\bibitem{Gumrukcuoglu:2011zh}
  A.~E.~Gumrukcuoglu, C.~Lin and S.~Mukohyama,
  JCAP {\bf 1203}, 006 (2012).
%
\bibitem{Gumrukcuoglu:2013nza}
  A.~E.~Gumrukcuoglu, K.~Hinterbichler, C.~Lin, S.~Mukohyama and M.~Trodden,
  Phys.\  Rev.\  D 88, {\bf 024023} (2013).



\end{thebibliography}
\end{document}